\documentclass[9pt,twocolumn,twoside,dvipsnames]{pnas-new}
\templatetype{pnasresearcharticle}

\setboolean{displaywatermark}{false}

\usepackage[T1]{fontenc}
\usepackage{microtype}
\usepackage{amsmath}
\usepackage{bm}
\usepackage{nicefrac}
\usepackage{graphicx}

\newcommand{\graphic}[2]{\includegraphics[width=#2\linewidth, type=pdf,ext=.pdf,read=.pdf]{#1}}
\newcommand{\dee}{\mathrm{d}}
\newcommand{\lcdm}{\Lambda\mathrm{CDM}}
\newcommand{\mpc}{\mathrm{Mpc}}
\newcommand{\msun}{\mathrm{M}_{\odot}}
\newcommand{\impc}{\mpc^{-1}}
\newcommand{\om}{\Omega_\mathrm{m}}
\newcommand{\ob}{\Omega_\mathrm{b}}
\newcommand{\se}{\sigma_8}
\newcommand{\ns}{n_\mathrm{s}}
\newcommand{\dm}{\delta_{\mathrm{m}}}

\title{Field Level Neural Network Emulator for Cosmological N-body Simulations}

\author[a,b,1]{Drew Jamieson}
\author[c,d,1]{Yin Li}
\author[c,e]{Renan Alves de Oliveira}
\author[f,g]{Francisco Villaescusa-Navarro}
\author[c,f]{Shirley Ho}
\author[g]{David N. Spergel}

\affil[a]{Max-Planck-Institut f\"ur Astrophysik, Karl-Schwarzschild-Straße 1, 85748 Garching, Germany}
\affil[b]{Department of Physics and Astronomy, Stony Brook University,
Stony Brook, NY 11794-3800, USA}
\affil[c]{Center for Computational Astrophysics,
Flatiron Institute, 162 5th Avenue, New York, NY 10010, USA}
\affil[d]{Center for Computational Mathematics,
Flatiron Institute, 162 5th Avenue, New York, NY 10010, USA}
\affil[e]{Centro de Ciências Exatas, Universidade Federal do Espírito
Santo. Av. Fernando Ferrari, 514. 29075-910. Vitória, ES, Brazil}
\affil[f]{Department of Astrophysical Sciences, 4 Ivy Lane, Princeton
University, Princeton, NJ 08544, USA}
\affil[g]{Simons Foundation, 160 5th Avenue, New York, NY 10010, USA}

\leadauthor{Jamieson \& Li}

\correspondingauthor{\textsuperscript{1}E-mail:
jamieson@mpa-garching.mpg.de \& eelregit@gmail.com}

\authorcontributions{Author contribution:
    DJ, YL, \& FVN designed research;
    DJ, YL, \& RAO performed research;
    DJ, YL, \& FVN contributed new reagents or analytic tools;
    DJ, YL, \& RAO analyzed data;
    DJ, YL, RAO, FVN, SH, \& DNS wrote the paper.
}

\authordeclaration{The authors declare no conflict of interest.}


\significancestatement{The distribution of galaxies and matter encodes crucial information about the laws and constituents of the Universe. Extracting this information requires accurate predictions of cosmic structure formation, usually obtained through expensive simulations. We develop a novel neural network emulator that reproduces cosmological simulation results with dramatic speed-up. Our model maps between early-universe conditions and late-time structure while capturing the correct dependence on cosmological parameters, making it a full-fledged emulator for large-scale structure. Conventional methods avoid running many simulations by focusing on summary statistics of observational data, which are prone to information loss. The success of our model enables more robust and complete analyses of large-scale structure observations without loss of information through data compression.}
\begin{abstract}
    We build a field level emulator for cosmic structure formation that is accurate in the nonlinear regime. Our emulator consists of two convolutional neural networks trained to output the nonlinear displacements and velocities of N-body simulation particles based on their linear inputs. Cosmology dependence is encoded in the form of style parameters at each layer of the neural network, enabling the emulator to effectively interpolate the outcomes of structure formation between different flat $\lcdm$ cosmologies over a wide range of background matter densities. The neural network architecture makes the model differentiable by construction, providing a powerful tool for fast field level inference. We test the accuracy of our method by considering several summary statistics, including the density power spectrum with and without redshift space distortions, the displacement power spectrum, the momentum power spectrum, the density bispectrum, halo abundances, and halo profiles with and without redshift space distortions. We compare these statistics from our emulator with the full N-body results, the COLA method, and a fiducial neural network with no cosmological dependence. We find our emulator gives accurate results down to scales of $k\sim 1\ \impc\, h$, representing a considerable improvement over both COLA and the fiducial neural network. We also demonstrate that our emulator generalizes well to initial conditions containing primordial non-Gaussianity, without the need for any additional style parameters or retraining.
\end{abstract}

\begin{document}
\maketitle

\thispagestyle{firststyle}

\ifthenelse{\boolean{shortarticle}}{\ifthenelse{\boolean{singlecolumn}}{\abscontentformatted}{\abscontent}}{}

% Introduction
\dropcap{W}e can learn more from the rapidly improving cosmological observations than we extract from traditional two-point statistics \cite{LSST:2008ijt,EUCLID:2011zbd,Dore:2014cca,Spergel:2015sza,DESI:2016fyo,DES:2017myr,DES:2021wwk}. However, utilizing this invaluable data set to its full potential and extracting the maximum amount physical information from its contents requires highly efficient and accurate methods of analysis and theoretical prediction. Recent advances in machine learning offer a potential path forward towards rapid and accurate methods of evolving the large-scale structure.

Standard practice in cosmology has been to employ summary statistics, such as power spectra and bispectra, in order to connect theory with observation \cite{2dFGRS:2001csf,SDSS:2003tbn,Beutler_2011,Blake_2012,BOSS:2016wmc,DES:2017myr,DAmico:2019fhj,Ivanov:2019pdj,eBOSS:2020yzd,DES:2021wwk,Philcox:2021kcw}. These statistics are also measured in configuration space, where they correspond to the two-point, three-point, and higher N-point correlation functions \cite{Slepian:2015hca, Philcox:2021hbm}. In this approach, a cutoff is straightforwardly imposed on small scales so as to only analyze those modes in the linear and quasilinear regimes, where accurate predictions can be made without reliance on numerical simulations. Another approach is to build summary statistics emulators based on simulations so that some information is extracted from nonlinear scales \cite{Chapman:2021hqe,Zennaro:2021bwy,Kobayashi:2021oud,Neveux:2022tuk,Zhai:2022yyk,Yuan:2022jqf}. However, the data compression involved in such analyses inevitably discards some information from observational surveys \cite{Charnock_2018,Samushia:2021ixs,Dai:2022dso}. More information may be extracted by considering higher N-point statistics beyond the power spectrum and bispectrum, but this requires computing a prohibitive number of mock data sets in order to determine the covariance matrices of these statistics. Novel summary statistics, including scattering wavelet transformations \cite{Cheng:2020qbx,Valogiannis:2021chp,Eickenberg:2022qvy,Valogiannis:2022xwu}, and k-nearest-neighbor distributions \cite{Wang:2021kbq} are being developed to decrease the information loss due to data compression while avoiding the computational challenges of higher N-point statistics. However, the goal of deriving robust constraints while maximizing information extraction from the data is best served by field level analysis.

Likelihood free inference at the field level exploits all of the information in the data. By extracting the maximum amount of information, field level analysis will achieve the tightest constraints on the values of cosmological parameters. Techniques along these lines include, deep learning models \cite{Ribli:2019wtw, Paco_2021a, Paco_2021b, Pablo_2022a}, forward modeling approaches \cite{Jasche_2013,Wang:2014hia,Jasche:2014kma,Ata:2014ssa,Seljak:2017rmr,Schmidt:2018bkr,Cabass:2020nwf}, and other methods of simulations based inference \cite{Alsing:2019xrx,Cranmer:2019eaq}. The most accurate theoretical predictions for structure formation at the field level come from expensive simulations. Even under the approximation where baryonic pressure and galaxy formation are neglected, so that all matter is treated as cold dark matter (CDM), state-of-the-art N-body simulations take a significant amount of time and computational resources, making them impractical for field level inference. These simulations evolve a large number of particles under the influence of Newtonian gravity in an expanding universe. Such algorithms transform a nearly uniform distribution of particles into the intricate structures of the cosmic web, a network of extremely dense dark matter halos and vast, nearly empty voids. This is a highly nonlinear process which is difficult to model accurately by other means.

Lagrangian Perturbation Theory (LPT) \cite{Scoccimarro:2001cj,Monaco:2001jg,Kitaura:2012tj,Chuang:2014vfa,Avila:2014nia,Stein:2018lrh}, lognormal generation \cite{Agrawal:2017khv}, particle mesh algorithms \cite{White:2013psd,Feng:2016yqz,Modi:2020dyb}, COLA \cite{Tassev2013, Howlett2015}, and neural network mock generation \cite{Berger:2018aey,Ramanah:2019cbm} all approximate nonlinear evolution with methods that are fast but either fail on small scales or do not accurately reproduce all of the key summary statistics. Another strategy is to develop a secondary algorithm responsible for correcting the small-scale errors of fast approximation schemes \cite{Kaushal:2021hqv}. This requires running multiple codes in succession, reducing gains in computational time and efficiency. Another approach that reduces the computational resources required for cosmological inference is the rescaling method, which converts the results from one simulation with a given set of cosmological parameters to the corresponding results for a different set of cosmological parameters \cite{Angulo2010, Contreras:2020kbv}. The rescaled results are specific to the set of initial conditions from the original simulation; the amplitudes of density modes are rescaled while the phases are held fixed. Thus the rescaling technique cannot be used to sample over both initial conditions and cosmological parameters simultaneously without running many N-body simulations.

In this work, we train a convolutional neural network (CNN) emulator capable of accurately reproducing the outcomes of N-body simulations for a wide range of cosmological parameter values. Unlike the previous methods mentioned above, our CNN based approach is both accurate in the deeply nonlinear regime and capable of computing the N-body evolution for arbitrary initial conditions. The nonlinear dynamics of structure formation involve couplings between all modes of the density field. Perturbative approaches typically approximate these couplings to second or third order. Our model learns the full, nonperturbative couplings and its dependency on $\om$.

Since our model is constructed from CNNs, it is fully differentiable, addressing the need to efficiently sample over both initial conditions and cosmological parameters simultaneously. This will greatly accelerate parameter inference from upcoming survey data. Our emulator is a significant advancement over the previous work in \cite{He:2018ggn}, where the authors trained a CNN to reproduce the FastPM displacements of $32^3$ particles. We also build on the previous work of \cite{AlvesdeOliveira:2020yix}, where the authors trained a CNN to reproduce the N-body displacement field for a fixed cosmology. In this work, we build the first field level, phase space emulator to predict the full N-body evolution of both the displacements and velocities of $512^3$ particles\footnote{The CNNs are trained on simulations with $512^3$ particles in a $1\ \mathrm{Gpc}\,h^{-1}$ box. This fixes the resolution that the CNNs can accurately model, but a different number of particles can be used by scaling the box size to keep the resolution fixed.} for a wide range of cosmological parameter values.

The emulator computes the N-body evolution on a flat, $\lcdm$ cosmological background, so the only relevant cosmological parameter is $\om$. Other $\lcdm$ parameters only affect the initial conditions (the model input), not the gravitational clustering, so they do not need to be included in the CNN design. One advantage of our method is that the CNN can generalize to extensions of $\lcdm$ when the additional cosmological parameters only affect the early universe, or initial conditions of the N-body simulation. As we will show, this includes primordial non-Gaussianity, which enters through the statistics of the linear field.

Our emulator can also easily be extended to include other cosmological parameters that do affect late-time, nonlinear clustering, such as the sum of neutrino masses, global spatial curvature, and parameters of dynamical dark energy models. In these cases, an extended model can be retrained with additional N-body simulations that include the relevant physics. Currently, our emulator makes predictions at redshift zero, but it is trivial to generalize this by including redshift as an extra parameter and training on simulation snapshots at earlier times.

\section*{Setup}
	\label{sec:setup}
	
    We train two CNN models to emulate the evolution of a system of N-body particles interacting under the influence of Newtonian gravity on an expanding cosmological background, which is specified by the matter fraction $\om$. One CNN takes the linear, also known at the Zel'dovich approximation (ZA), displacement field at redshift $z=0$ and the value of $\om$ as inputs and outputs the nonlinear displacement field at redshift $z=0$. The other CNN takes the ZA velocity field at redshift $z=0$ and the value of $\om$ as input and outputs the nonlinear velocity field at redshift $z=0$. Initially, the N-body particles are distributed on a uniform grid with positions $\mathbf{q}$, and their final positions at redshift zero are
    \begin{align}
        \label{eq:dis}
        \mathbf{x}(\mathbf{q}) = \mathbf{q} + \mathbf{\Psi}(\mathbf{q}) \, ,
    \end{align}
    where $\mathbf{\Psi}(\mathbf{q})$ is the final displacement for the particle initially at grid site $\mathbf{q}$. The final velocities of each particle are,
    \begin{align}
        \mathbf{v}(\mathbf{q}) \equiv \dot{\mathbf{x}}(\mathbf{q}) = \dot{\mathbf{\Psi}}(\mathbf{q}) \, ,
    \end{align}
    where the dots denote time derivatives. Since all of the particles have the same mass, $M_p$, the velocities are really mass weighted and represent the momentum field,
    \begin{align}
        \label{eq:lmom}
        \mathbf{p}(\mathbf{q}) = M_p\, \mathbf{v}(\mathbf{q}) \, .
    \end{align}
    Our CNN model effectively nonlinearizes the linear inputs, directly approximating the outcome of an N-body simulation by learning the mode couplings of the dark matter field.
 
	We train our CNN emulator with 2000 simulations from the Quijote latin hypercube N-body suite \cite{Villaescusa-Navarro:2019bje}. Each has a unique set of five cosmological parameters: $\om$, $\ob$, $\se$ $\ns$, and $h$. The values of these cosmological parameters are randomly sampled in a five dimensional latin hypercube. Additionally, each simulation has a unique random seed for its initial condition, so all of the initial conditions are different in every simulation. The simulations were run using the N-body code Gadget-3 \cite{Springel:2005mi} with $512^3$ particles in a $1~\mathrm{Gpc}\,h^{-1}$ box. Our CNNs are trained on the redshift zero simulation snapshots.

	We split the 2000 latin hypercube samples into a training set of 1757 cosmologies, a validation set of 122 cosmologies and a testing data set of 121 cosmologies. The loss on the training data set is used to determine back propagation gradients through the neutral network and to update the model parameters during each training epoch. The loss on the validation set is monitored as a diagnostic for overfitting during training. After training the emulator, the testing simulations are used to determine how the model performs on data independent from the training and validation sets.

	In addition to the usual convolutional parameters of a CNN, our emulator contains extra parameters at each layer that encode the $\om$ dependence of clustering across multiple scales. The value of $\om$ is fed to each layer of the CNN and transformed to an array with the same dimension as the layer's input. The convolution kernels are multiplied by this array prior to performing the convolution. We described this in greater detail in the materials and methods section. Following StyleGAN2 \cite{StyleGAN2}, we refer to $\om$ as a \emph{style} parameter, and we refer to the emulator as the \emph{styled neural network} (SNN) model. We compare this model against two competitors. The first has the same neural network architecture except it does not include the style parameter, so it has no cosmology dependence. We refer to this model as the \emph{fiducial neural network} (FNN) model, since it was trained on a set of N-body simulations in the Quijote simulation suite with only the fiducial values of the five cosmological parameters. This comparison allows us to evaluate the gains in accuracy achieved by including the style parameter, which gives the SNN its cosmology dependence. 

	The second competitor we consider is COLA \cite{Tassev2013}, implemented with L-PICOLA \cite{Howlett2015}. The COLA method is a fast approximation for solving the N-body equations of motion using tens, rather than thousands of integration time steps. Although the COLA method is fast, it produces inaccurate results on small scales. Comparing with COLA gives us a basis for assessing the gains in accuracy from implementing a CNN to predict the outcomes of simulations rather than using approximate N-body solvers. Below we present results from the predictions of our SNN emulator, and compare the with the full N-body simulations, the FNN, and COLA.
	
\section*{Results}
	\label{sec:resu}
	
	\begin{figure*}
		\centering
		\graphic{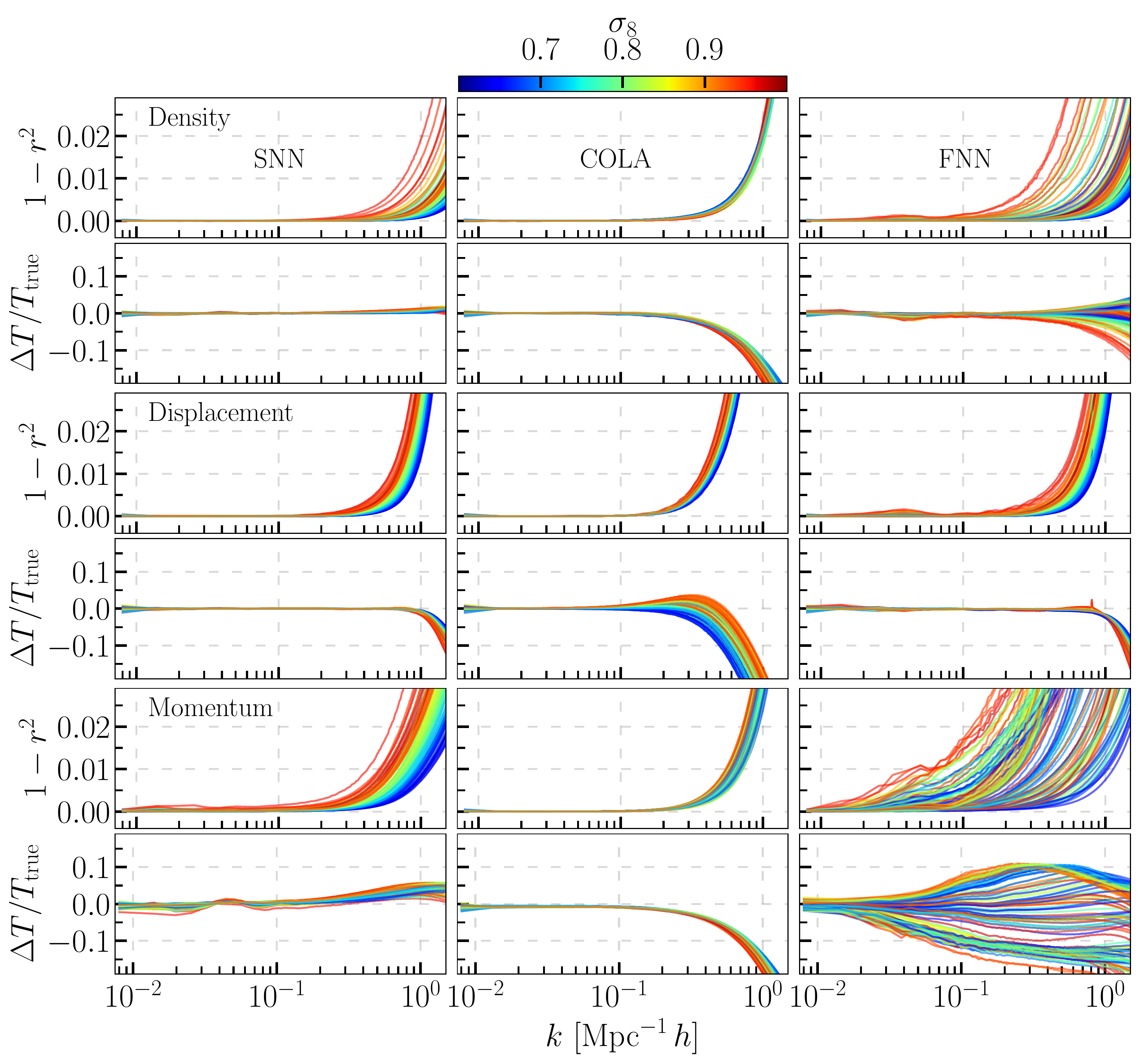}{0.63}
		\caption{Power spectra errors for the styled neural network (left), COLA (middle), and the fiducial neural network (right). Each pair of rows shows the stochasticities and transfer function errors for the Eulerian density field power spectra (top two), the Lagrangian displacement field power spectra (middle two), and the Eulerian momentum field (bottom two) power spectra. The color of each curve corresponds to its value of $\se$ according to the color bar at the top.}
		\label{fig:Pe}
	\end{figure*}
		
    While the main goal of building a field level emulator is to do full field level inference, it is useful to assess the model's accuracy by evaluating summary statistics. The emulator can also be used for fast mock generation to study the covariance of summary statistics, for which we need to validate its accuracy. In this section we compare the power spectra (two-point statistics) of the matter density modes, the N-body displacement modes, and the momentum field modes from the full N-body simulations with our emulator and its competitors. We then compare the bispectrum (three-point statistics) of the matter field modes. In addition to these N-point statistics, we demonstrate our emulator reproduces the abundances of dark matter halos and their profiles better than its competitors. We then determine the effects of redshift space distortions (RSD) on some of these observables and how they compare with the full N-body results. 

	\subsection*{Density Power Spectra}
		\label{ssec:denpk}

		For each simulation in the testing data set we estimate the Eulerian density field using a cloud-in-cell (CIC) particle mesh assignment on a grid with $1024^3$ voxels. At the voxel centered on position $\mathbf{x}$, we sum the CIC particle weights to obtain the effective number of particles $n(\mathbf{x})$, and then determine the overdensity $\dm(\mathbf{x}) = n(\mathbf{x}) / \bar{n} -1$ with respect to the mean number of particles per cell $\bar{n}$. We Fourier transform the density fields using FFTW3 \cite{Frigo:2005zln} and deconvolve the CIC window function, yielding the density modes $\dm(\mathbf{k})$. We then compute the density power spectrum $P_{\mathrm{mm}}(k)$, given by
		\begin{align}
			\langle \dm(\mathbf{k}) \dm(\mathbf{k}') \rangle = (2\pi)^3 \delta^{3}_{\mathrm{D}}(\mathbf{k} + \mathbf{k}') P_{\mathrm{mm}}(k) \, .
		\end{align}
		Here $\delta^{(3)}_{\mathrm{D}}(\mathbf{k})$ denotes the 3D Dirac delta function, which enforces the homogeneity of the density statistics, while the fact that the power spectrum depends only on the magnitude $k \equiv |\mathbf{k}|$ is required by isotropy. In practice, we compute the power spectrum as the mean squared mode amplitude in bins the width of the Nyquist wave number of the Fourier mesh.

        Let $\delta_{\mathrm{m,pred}}$ be the predicted density field constructed from either the emulator, COLA, or the fiducial model, and let $\delta_{\mathrm{m,true}}$ be the density field constructed from the N-body simulations. We characterize the errors in the model predictions using the cross correlation coefficient,
		\begin{align}
			\label{eq:sto}
			r(k) = \frac{ \langle \delta_\mathrm{\mathrm{m,pred}}(\mathbf{k}) \delta_{\mathrm{m,true}} (\mathbf{k'}) \rangle } {\sqrt{\langle \delta_{\mathrm{m,pred}}(\mathbf{k}) \delta_{\mathrm{m,pred}}(\mathbf{k'}) \rangle \langle \delta_{\mathrm{m,true}}(\mathbf{k})\delta_{\mathrm{m,true}}(\mathbf{k'}) \rangle }} \, .
		\end{align}
		The stochasticity is defined as $1 - r(k)^2$, which quantifies the excess fraction of correlation in the prediction that cannot be accounted for in the target simulation data. We also characterize the errors using the fractional difference between the predicted and true transfer functions,
		\begin{align}
			\label{eq:tfe}
			\frac{T_\mathrm{pred}(k)}{T_\mathrm{true}(k)} -1 = \sqrt{\frac{P_{\mathrm{mm},\mathrm{pred}}(k)}{P_{\mathrm{mm},\mathrm{true}}(k)}} - 1 \, .
		\end{align}
		The results for these density power spectrum errors are shown in the top two rows of Fig. \ref{fig:Pe}. The color of each curve indicates the value of $\se$ from the corresponding simulation. We use $\se$ rather than the style parameter $\om$ to label the curves because the small-scale errors are determined by the abundances of high-density structures; $\se$ parameterizes the amplitude of small-scale power, so higher $\se$ generally indicates a greater abundance of collapsed objects (i.e. dark matter halos).
		
		The SNN is considerably more accurate than COLA at scales out to $k\sim 1\ \impc\, h$. The emulator achieves percent level accuracy in the transfer functions for all cosmologies. The emulator also achieves percent level stochasticities for all cosmologies except those with extremely high $\se$. As we will continue to see in results presented later, it is the virialized motion inside of collapsed regions that limit the emulator's accuracy, and this explains the elevated errors in the matter power spectrum for high $\se$ simulations. Due to the fast orbits and fairly chaotic behavior of particles inside a virialized halo, if we output the simulation snapshots at slightly earlier or later times the particle positions and velocities can change significantly. Since the CNNs map the ZA inputs to nonlinear fields at the particle level, it is very difficult for the CNNs to learn general rules about the fates of these particles. For COLA the accuracy is limited by the combination of LPT and large time steps. Even with these limitations, in the worst cases the emulator achieves stochasticities of less than 2\% at $k=1~\impc\,h$ which is comparable with COLA's accuracy. We also see that the FNN errors have a much stronger cosmology dependence, so the SNN is using its style parameter to effectively interpolate between different cosmological models and make more accurate predictions.
		
		The emulator performs worst on cosmologies with a combination of high $\se$ and low $\om$. The large-scale power in these cosmologies has an extremely high amplitude. This results in a large divergence in the linear displacement field on large scales, so particles generally flow out further from their initial positions. In these extreme cosmologies, the nonlinear displacements are more nonlocal, sourced by a wider region of the density field. This poses a different limitation on the SNN, besides small-scale resolution, since it has a limited field of view for the environment surrounding each particle. In principle this limitation can be alleviated by adding more convolutional layers, widening the model's field of view. However, in practice this would increase the evaluation time and require more memory for the large number of additional parameters. We initially implemented the neural network with only three layers, and in this case the sensitivity of the stochasticities to low values of $\om$ was even more dramatic and the worst cases were worse than COLA. Increasing to four layers improved these worst cases without too severe a cost in memory and computational time.

	\subsection*{Displacement Power Spectra}
		\label{ssec:dispk}

        From Eq. (\ref{eq:dis}), the N-body displacement field is defined with respect to the initial Lagrangian grid, so these can be Fourier transformed directly. The displacement power spectrum, $P_{\mathbf{\Psi\Psi}}(k)$, is then given by
		\begin{align}
			\langle \mathbf{\Psi}(\mathbf{k}) \cdot \mathbf{\Psi}(\mathbf{k'}) \rangle = (2\pi)^3 \delta^{(3)}_{\mathrm{D}}(\mathbf{k} + \mathbf{k'}) P_{\mathbf{\Psi\Psi}}(k) \, ,
		\end{align}
		where $\mathbf{\Psi}(\mathbf{k})$ are the Fourier modes of the displacement field and the dot indicates taking the Cartesian inner product between the displacement mode vectors. The displacement stochasticity and transfer function errors are defined analogously to those of the density field in Eqs. (\ref{eq:sto}--\ref{eq:tfe}). These errors are plotted in the middle two rows of Fig. \ref{fig:Pe}.
		
		Again, we find that the SNN achieves smaller errors than COLA, and has less cosmology dependent errors than the FNN. Similarly to the density power spectrum results, the worst cases for the emulator are the high $\se$ cosmology with low $\om$. The SNN displacement stochasticities are less than 3\% at $k=1~\impc\,h$ and the transfer function errors are negligible ($<0.1$\%) down to this scale.
		
		Note that the small bump in the FNN errors at $k\sim0.04\ \impc/h$ corresponds to the baryonic acoustic oscillations (BAO) in the power spectrum. The shape of these oscillations depends on the ratio $\ob/\om$. Since the FNN only encountered one example of this ratio, it is unable preserve the shape of the BAO oscillations in cosmologies with parameters that are far from the fiducial values of its training data. A similar bump in the FNN errors can also be seen in the results for the density power spectrum.

	\subsection*{Momentum Power Spectra}
		\label{ssec:mompk}

        The previous two power spectra depended on the displacement field alone. To compare the accuracy of the velocity part of our model, we construct the momentum power spectrum. We estimate the momentum field, $\mathbf{p}(\mathbf{x})$, by distributing the particles to a $1024^3$ mesh using the CIC assignment scheme. Note that this is an Eulerian momentum field, not the Lagrangian momentum field in Eq. (\ref{eq:lmom}), which was defined with respect to the initial particle grid. We Fourier transform the momentum field and deconvolve the CIC window function to obtain the momentum modes $\mathbf{p}(\mathbf{k})$. We then compute the momentum power spectrum $P_{\mathbf{pp}}(k)$,
		\begin{align}
			\langle \mathbf{p}(\mathbf{k}) \cdot \mathbf{p}(\mathbf{k'}) \rangle = (2\pi)^3 \delta^{(3)}_{\mathrm{D}}(\mathbf{k} + \mathbf{k'}) P_{\mathbf{pp}}(k) \, ,
		\end{align}
		using the same method that was used for the density power spectrum. The errors are plotted in the bottom two rows of Fig. \ref{fig:Pe}.
		
		The SNN momentum transfer functions are much more accurate on small scales than those from COLA, and show much less cosmology dependence than the FNN transfer function errors. The same is true for the stochasticities for all but the most extreme, high $\se$ low $\om$ cosmology, where the stochasticities are slightly larger, although comparable with COLA. The emulator has stochasticities of less than 2\% at $k=1\impc\,h$ for most cosmologies. The three worst cases have stochasticities of 4--5\% at this scale. The SNN transfer functions at $k=1\impc\,h$ are biased slightly too high by a few percent, but remain considerably more accurate than COLA.
		
		In general, the SNN momentum power spectrum errors are worse than those from the density or displacement power spectra. This is due to the fact that constructing the momentum field relies on the particle displacements, which are taken from the output of the displacement SNN. Thus the errors from the displacement field propagate to the momentum field.

		\begin{figure*}
			\centering
			\graphic{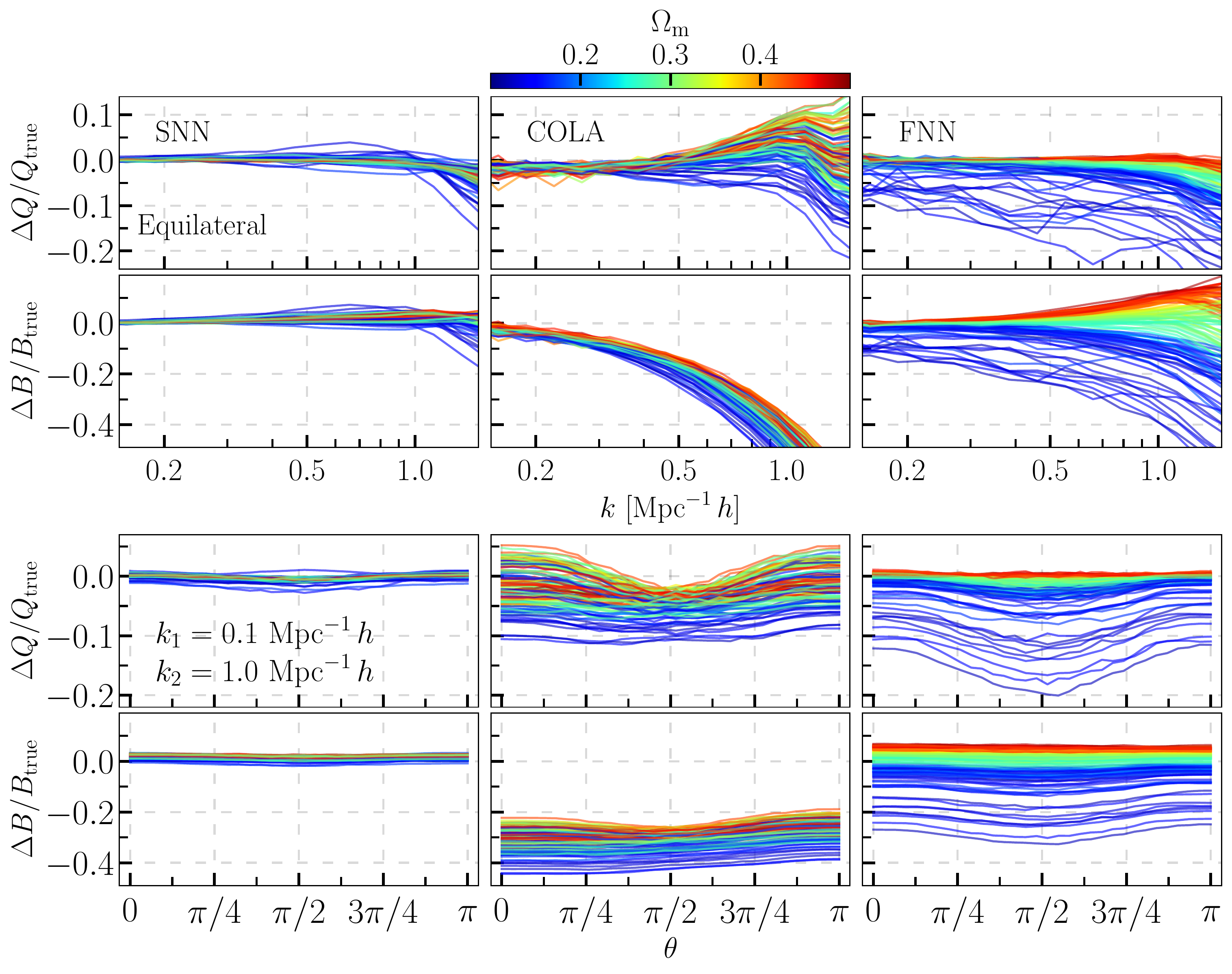}{0.7}
			\caption{Bispectrum errors for equilateral configurations (top two rows) and configurations obtained from varying the angle between two wave vectors with fixed magnitudes (bottom two rows). The bottom rows in each pair show the errors in the bispectrum, while the top rows show the errors in the reduced bispectrum. The color of each curve corresponds to its value of $\om$, according to the color bar at the top.}
			\label{fig:Qe}
		\end{figure*}

	\subsection*{Density Bispectra}
		\label{ssec:denbk}

		In addition to the two-point statistics we also considered the three-point statistics of the Eulerian matter density field. The matter bispectrum $B(k_1, k_2, k_3)$ is defined,
		\begin{align}
			\langle \dm(\mathbf{k}_1) \dm(\mathbf{k}_2) \dm(\mathbf{k}_3) \rangle = \ & (2\pi)^3 \delta^{(3)}_{\mathrm{D}}(\mathbf{k}_1 + \mathbf{k}_2 + \mathbf{k}_3) \nonumber \\ & \times B(k_1, k_2, k_3) \, .
		\end{align}
		Since the Dirac delta function requires the three wave vectors to form a closed triangle, the bispectrum can also be expressed as a function of the magnitude of two wave vectors and the angle between them, $B(k_1, k_2, \theta)$. We use both notations interchangeably. It is also convenient to define the reduced bispectrum,
		\begin{align}
			Q(k_1, k_2, k_3) \equiv \frac{B(k_1, k_2, k_3)}{P_1 P_2 + P_2 P_3 + P_3 P_1} \, ,
		\end{align}
		where $P_i = P_{\mathrm{mm}}(k_i)$ is the density power spectrum evaluated at $k_i$. We use the P\textsc{ylians}3\footnote{\url{https://github.com/franciscovillaescusa/Pylians3}} library to compute the bispectrum.
		
		We characterize the bispectrum errors by taking the fractional difference with respect to the N-body bispectrum. These are shown in Fig. \ref{fig:Qe} for two different configurations. The two rows of plots show the bispectra in the equilateral configuration as a function of the magnitude of the three wave vectors. The bottom two rows show the bispectra fixing $k_1 = 0.1\ \impc\,h$ and $k_2 = 1\ \impc\,h$ as a function of the angle between them. The top row of each set of plots shows the reduced bispectrum errors while the bottom row shows the error in the bispectrum. The color of each bispectrum curve corresponds to its value of $\om$. 
		
		Unlike the power spectrum errors, the SNN bispectrum errors do not exhibit a strong dependence on $\om$ or $\se$. The SNN bispectrum errors are significantly better than those from COLA or the FNN. This demonstrates that the model is inferring general physical principles about nonlinear gravitational clustering and its cosmology dependence, enabling it to accurately model structure formation on small scales. The SNN errors are less than 5\% at $k=0.5\ \impc\,h$ and do not increase significantly until after $k=1\ \impc\,h$. The errors in both the bispectra and reduced bispectra are comparable for the SNN because its power spectra errors are small on these scales.

	\subsection*{Halos}
		\label{ssec:halos}
		
		\begin{figure*}
			\centering
			\graphic{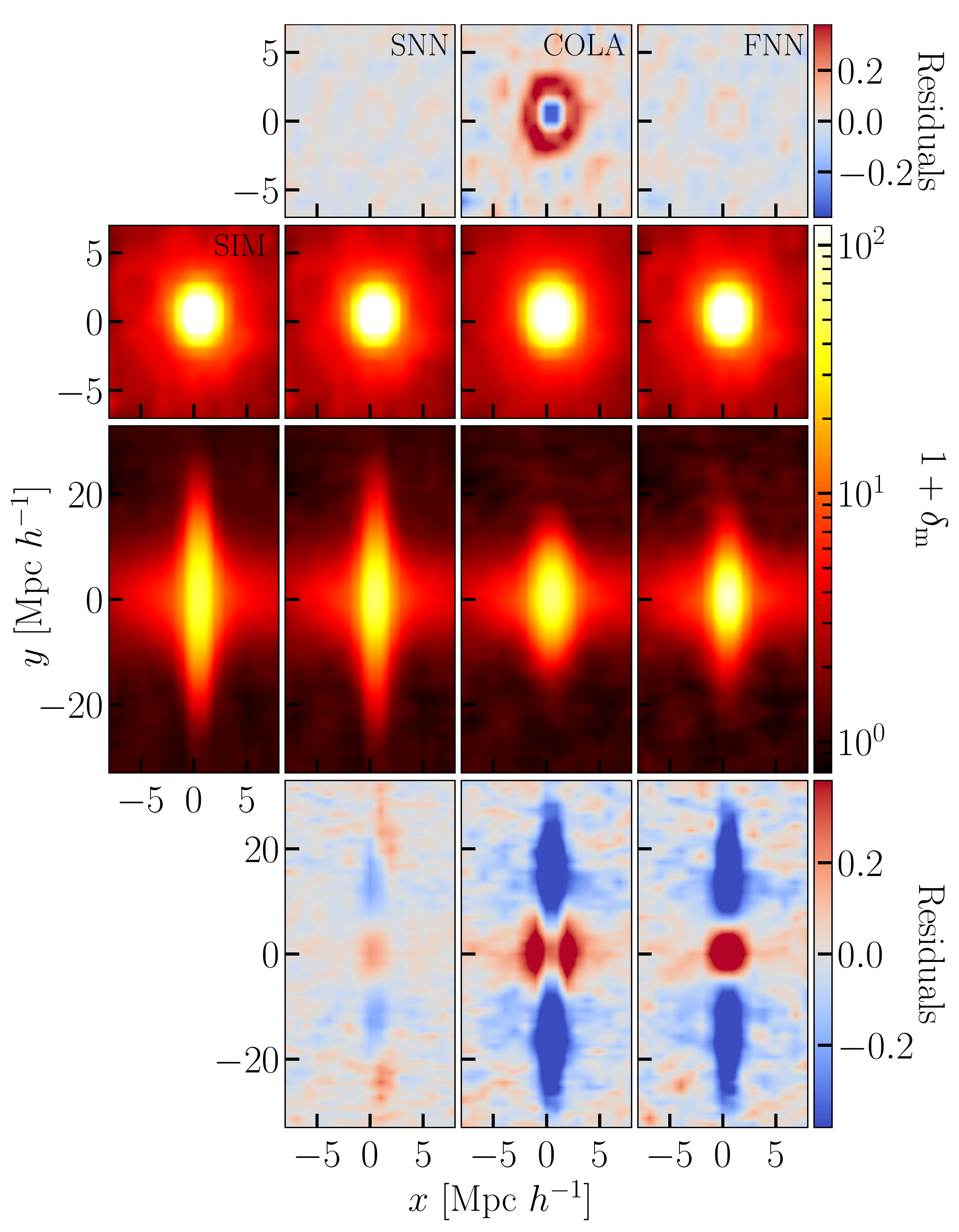}{0.44}
			\graphic{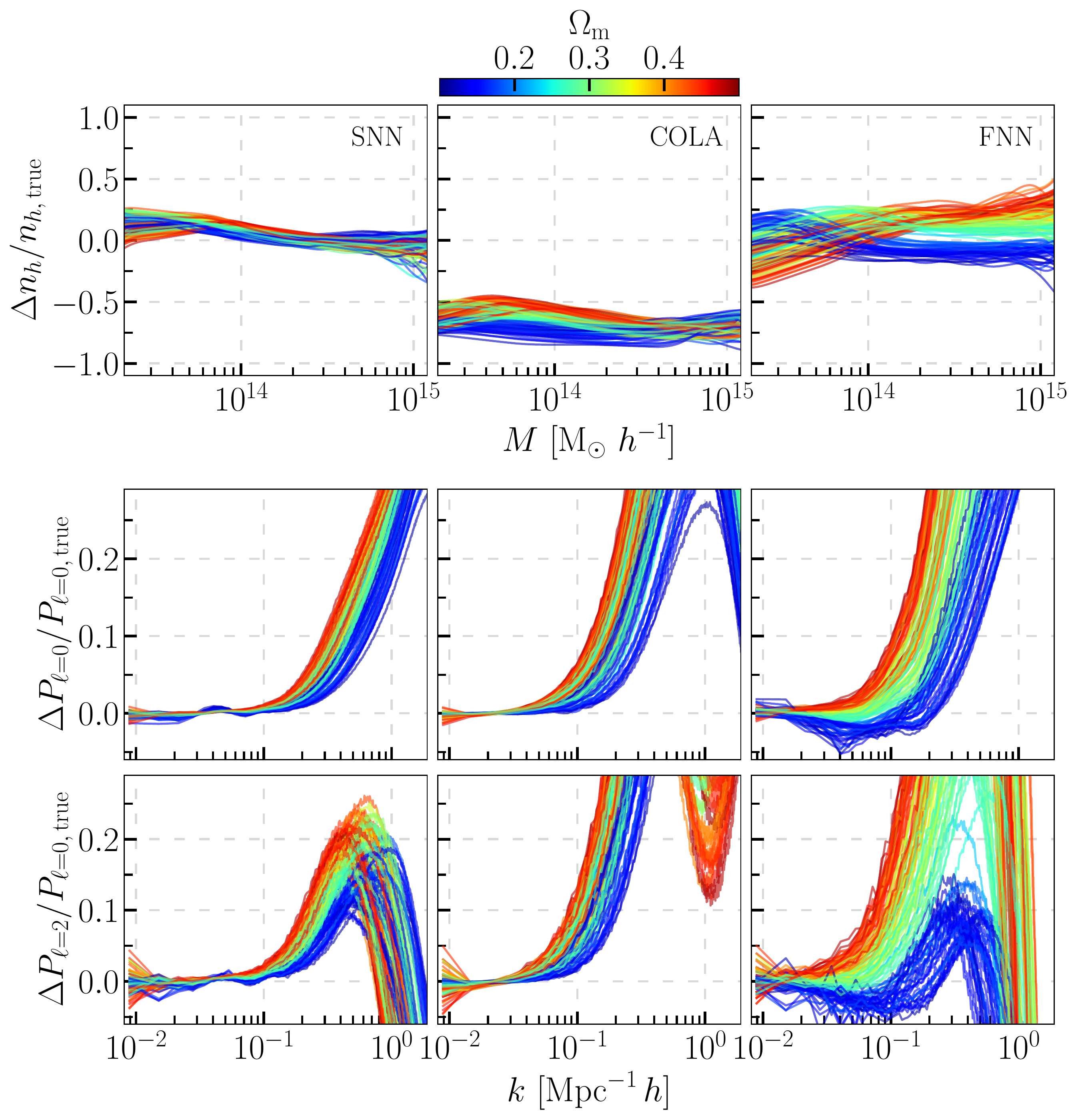}{0.55}
			\caption{Left: stacked profiles of 500 halos in real space (top two rows) and redshift space (bottom two rows) with mean mass of $5\times10^{14}\ \msun\,h^{-1}$. The top and bottom rows show the residuals with respect to the N-body simulation. The middle two rows show the profiles. Right: halo mass function errors (top) and errors in the multipole moments of the redshift space matter power spectrum (bottom two rows). The color of each curve corresponds to its value of $\om$, according to the color bar at the top.}
			\label{fig:stk}
		\end{figure*}

        The matter density field is not directly measured in large-scale structure surveys. Instead, biased tracers of the underlying density field, such as galaxies, are used to infer the statistics and initial conditions of the density field. Under gravitational clustering, the dark matter forms high-density, virialized structures, called halos, which are the environments in which galaxies form and reside.

		We identify halos in the N-body, CNN, and COLA outputs using the halo finder code Rockstar \cite{Behroozi:2011ju}. Rockstar identifies halos using a friend-of-friends algorithm and then determines halo properties by spherical overdensity calculations out to each halo's virial radius. Particles that are not bound to a halo's center of mass are omitted, so the six dimensional phase space of the N-body particle data is utilized by this algorithm. Rockstar provides us with an excellent framework for further assessing the combined, full phase phase predictions of our emulator.
		
		One important observable related to dark matter halos is their density profiles. These density profiles are the result of extremely nonlinear processes of gravitation collapse, mergers, and accretion. To construct halo density profiles from our data, we select a sample of 500 halos with an average mass of $5\times10^{14}\msun /h$. These were taken from a single simulation in the testing data set with cosmological parameters closest to the fiducial $\lcdm$ cosmology. We construct the density field from the positions of the N-body particles using a CIC mesh assignment on a grid of $0.5~\mpc\, h^{-1}$ in cell length. We average the densities of each halo in the $z$-direction over a $3\ \mpc\, h^{-1}$ slice centered on each halo's center of mass. The results are shown in the top two rows of the upper left of Fig. \ref{fig:stk}.

        The SNN achieves much more accurate halo interiors than COLA, and also shows improvements over COLA in the environments surrounding halos. The SNN even shows small improvements over the FNN, even though these halos form in a cosmology close to the FNN's training data. This demonstrates that even small changes in $\om$ can result in a loss of accuracy for the FNN in the deeply nonlinear regime.

        Another important observable related to halos is their number density as a function of mass, $n_h(M)$, which is known as the halo mass function. We construct the halo mass function for each testing cosmology and compute its error as the fractional difference with respect to the N-body results. These errors are shown in the top row of plots in the upper right of Fig. \ref{fig:stk}.

        The SNN shows a dramatic improvement over COLA, and again we find reduced cosmology dependence in the errors compared with the FNN. The abundances of low-mass halos ($M<5\times10^{13}\,\msun/h$) from the SNN are biased high by about 10--20\%, but are scattered around the N-body mass functions at higher mass. To be fair to COLA, its purpose is not to accurately recover the small-scale details of halos, and its systematically low halo abundances are well known. Abundance matching methods have been developed to identify dense structures in COLA output that correspond to halos from the full N-body evolution \cite{Izard:2015dja}.

	\subsection*{Redshift Space Distortions}
		\label{ssec:rsd}

        In large-scale structure surveys, the distance to a galaxy is inferred through its observed redshift. This redshift is not due to cosmic expansion alone, but gets an additional doppler contribution from the galaxy's peculiar velocity. Even with the correct cosmological model, the observed positions of galaxies are radially displaced compared to their true positions. The shapes of galaxies are also distorted along the line of sight. These effects are known as redshift space distortions (RSD). Matter inside of halos tends to have large, virialized velocities, which results in a smearing of the halo density along the line of sight, known as the Fingers of God phenomenon. On larger scales, where matter falls towards the local minimum of the gravitational potential, spherical profiles are squashed along the line of sight. It is crucial to accurately describe the late-time, nonlinear density field in redshift space in order to compare with realistic survey data.

		We use the P\textsc{ylians}3 library to apply RSD to the N-body particle data, and then construct the density profiles for halos in redshift space. The stacked redshift space profiles of the same 500 halos as in the previous subsection are shown in the bottom left corner of Fig. \ref{fig:stk}, along with their residuals with respect to the N-body simulations. All of the models exhibit the larger scale flattening effect due to coherent infall, which appears as the two bulges on either side of the halo. However, only the SNN accurately reproduces the scale of the Fingers of God smearing along the $y$-axis compared with the N-body simulations.

        As a final test of our model's accuracy in the nonlinear regime, we consider the multipole moments of the matter power spectrum in redshift space. Since RSDs contribute only radial distortions, they select the radial direction as special and induce anisotropy in the power spectrum. This anisotropy can be characterized by expanding the redshift space power in multipole moments,
        \begin{align}
            P_{\mathrm{mm}}(\mathbf{k}) = \sum_{\ell} (2\ell + 1) P_{\ell}(k) \mathcal{L}_{l}\big(\cos(\theta)\big) \, .
        \end{align}
        Here $\mathcal{L}_{\ell}$ are the Legendre polynomials and $\theta$ is the angle between the wave vector $\mathbf{k}$ and the radial line of sight direction. We used the Pylians library to determine the monopole ($\ell = 0$) and quadrupole ($\ell = 2$) moments of the redshift space density power spectrum. The fractional errors with respect to the N-body results are shown in the bottom right of Fig. \ref{fig:stk}

        The SNN reproduces the monopole and quadrupole power spectra more accurately than COLA on small scales. Once again, we see the SNN errors have less cosmology dependence than the FNN errors, indicating that the full phase space evolution is modeled better across a wide range of $\om$ values due to its inclusion as a style parameter. Note that since RSDs tend to move small-scale power to larger scales, the errors are larger at the $1\ \impc\,h$ scale compared with the power spectrum errors in real space.

 	\subsection*{Primordial non-Gaussianity}
 		\label{ssec:fnl}
 		
 		\begin{figure*}
			\centering
			\graphic{./Figures/fnl_errors}{0.42}
			\graphic{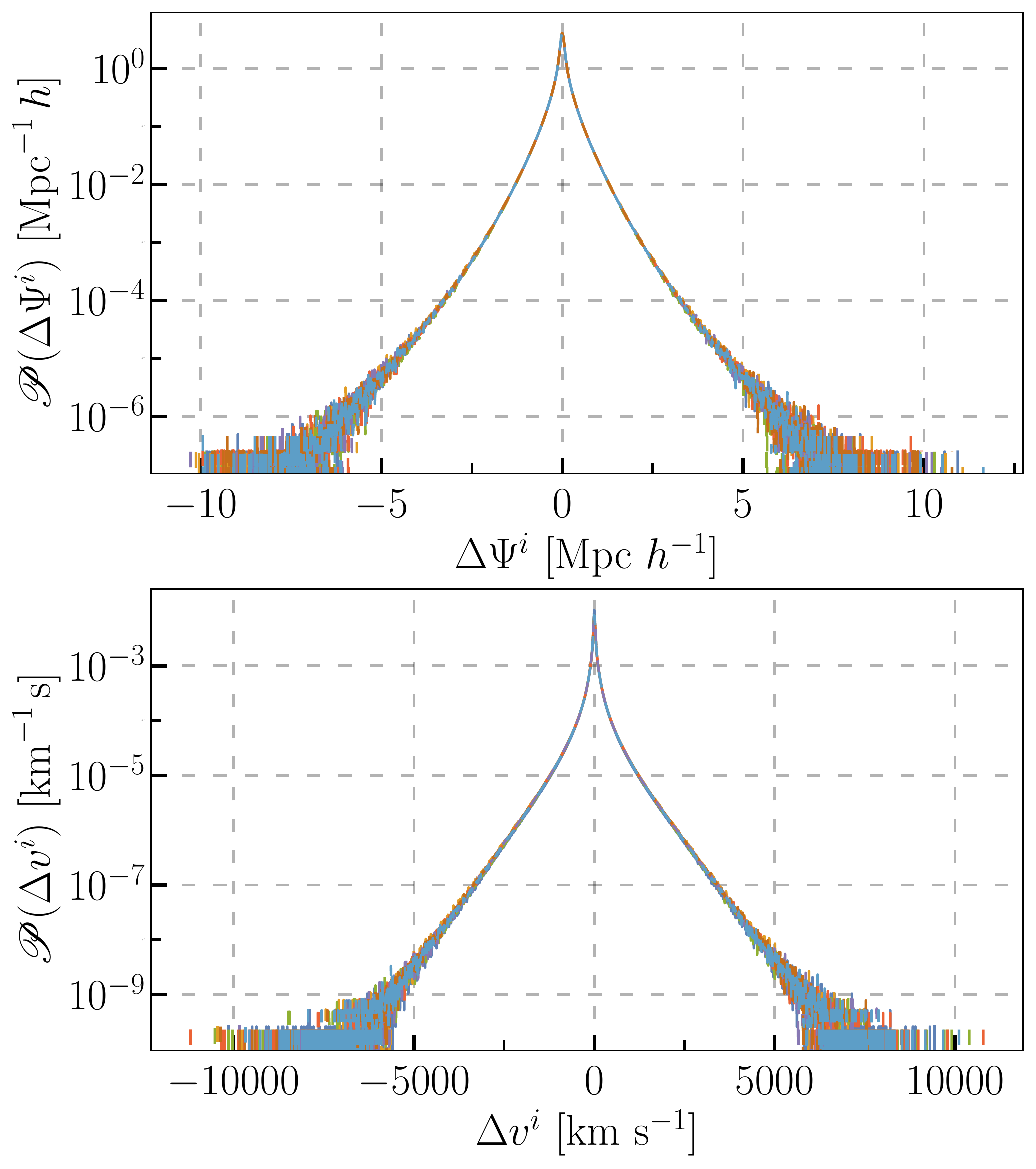}{0.44}
			\caption{Left: Standard deviation in SNN errors of the displacement field (top) and velocity field (bottom) for Gaussian (G) and PNG simulations, including local (LC), orthogonal (OR), and equilateral (EQ) primordial bispectrum templates. The $\pm$ signs indicate the sign of $f_{\mathrm{NL}}$, which has a magnitude $|f_{\mathrm{NL}}|=100$ for all PNG simulations. Right: SNN component error probability density functions for displacements (top) and velocities (bottom) from Gaussian and PNG simulations. No labels or legend are shown since the distributions are indistinguishable.}
			\label{fig:fnl}
		\end{figure*}
 		
        Searching for observational evidence of primordial non-Gaussianty (PNG) is one of the highest priority objectives of contemporary large-scale structure surveys. If detected, PNG would provide crucial evidence for inflation and possibly rule out a wide range of early universe models \cite{Creminelli:2004yq,Meerburg:2019qqi}. The simulations that our emulator was trained on have purely Gaussian initial conditions. Non-Gaussianity can be added to the initial conditions by first drawing the modes of a random Gaussian scalar potential $\Phi_{\mathrm{g}}(\mathbf{k})$, and then constructing the non-Gaussian potential \cite{Scoccimarro:2011pz},
        \begin{align}
            \Phi_{\mathrm{ng}}(\mathbf{k}) = \Phi_{\mathrm{g}}(\mathbf{k}) + f_{\mathrm{NL}} \!\int\! \frac{\dee^3k'}{(2\pi)^3} K(\mathbf{k}', \mathbf{k} - \mathbf{k}') \Phi_{\mathrm{g}}(\mathbf{k}') \Phi_{\mathrm{g}}(\mathbf{k} - \mathbf{k}') \, .
        \end{align}
        Here $f_{\mathrm{NL}}$ sets the amplitude of the primordial bispectrum, and $K(\mathbf{k}_1, \mathbf{k}_2)$ is an integration kernel specifying the shape of the primordial bispectrum. There are several typical choices for this kernel, usually referred to as bispectrum templates \cite{WMAP:2003xez,Babich:2004gb,Senatore:2009gt}. These include the \emph{local} (LC), \emph{orthogonal} (OR), and the \emph{equilateral} (EQ) templates. Since PNG enters only through the initial conditions of a simulation, our model should immediately generalize to these cosmologies without the need for additional parameters or training.

        To test this, we use the new QUIJOTE-PNG \citep{Coulton:2022qbc} N-body simulations. These augment the previous QUIJOTE data set with simulations that include the PNG templates mentioned above in their initial conditions. We use a set of seven simulations, one with Gaussian (G) initial conditions and two from each of the bispectrum template types listed earlier, setting $f_{\rm NL} = \pm 100$. All other cosmological parameters are the same as in the fiducial cosmology and the simulation configurations are identical to those of the training data. We compute the ZA displacements and velocities at redshift zero for all of these simulations, and predict the nonlinear configurations using our emulator. We then compare these with the full N-body simulations. The results are shown in Fig. \ref{fig:fnl}.

        The plot on the upper left shows the standard deviation of displacement errors for the Gaussian and non-Gaussian simulations. There is no significant or systematic increase in error in the simulations with PNG, some even have smaller errors than the Gaussian case. The bottom left plot shows the standard deviation of velocity errors, again without any systematic increase from the presence of PNG in the initial conditions. The two plots on the right show the probability density functions for errors in the displacement components (top) and the velocity components (bottom) for all seven simulations. We do not include a legend to label the different template types, since their error distributions are indistinguishable. This indicates that the emulator generalizes to these PNG simulations as expected. 

        These results are unsurprising, since the model has been exposed to an enormous diversity of density environments during training. The presence of PNG alters the statistics of these configurations, but the physics of nonlinear gravitational clustering in these environments remains the same. It is this underlying physics of late-time structure formation that the emulator infers from its training data. Although further tests are still required to determine how well the emulator preserves observational features of PNG in various summary statistics, including the power spectrum and bispectrum, as well as at the field level. We leave this to future work.
 		
\section*{Conclusion}
	\label{sec:conc}
	
	We have trained and tested a full phase space emulator for N-body simulations based on two styled neural networks. The emulator achieves percent level accuracy on deeply nonlinear scales of $k\sim1\ \impc\, h^{-1}$. We have demonstrated this for the density power spectrum, the Lagrangian displacement power spectrum, and the momentum power spectrum. The emulator also achieves accuracy on the order of a few percent for the density bispectrum. The SNN model reproduces abundances of dark matter halos with an accuracy on the order of 10\%. We have also demonstrated that the emulator accurately predicts the full phase space N-body evolution, reproducing the features of RSDs in halo profiles and the matter density power spectrum.
	
	The accuracy of our emulator represents a significant improvement over other fast simulation techniques, such as the COLA method. Comparisons against a neural network model without the style parameter, thus lacking cosmology dependence, demonstrate that the emulator has learned to interpolate between cosmologies within the range of its training data. The emulator was trained on simulations with flat, $\lcdm$ cosmological backgrounds. Due to the design choice of predicting residuals between the full nonlinear displacements and velocities and their linear approximations, the emulator generalizes well to extensions to $\lcdm$ that only affect the early universe, and thus enter only through the linear fields. We have demonstrated this for the case of primordial non-Gaussianity.
	
	For other extensions to $\lcdm$ that do affect late-time clustering, and also for redshift dependence, our emulator can easily be augmented without the need for significant changes to the underlying code. Additional style parameters can be defined, and then the model can be retrained on simulation data that includes the new physics. This may only require a small amount of new training data and a small amount of retraining time if transfer learning techniques are employed to retain the current capabilities of the emulator.
	
	The accuracy, speed, and differentiability of our emulator makes it an ideal tool for field level, forward modeling approaches to cosmological inference. Additionally, the emulator can efficiently generate large numbers of mocks for studying covariance matrices of various observables. Our emulator demonstrates the power of machine learning techniques to accelerate onerous computational tasks, which will help to achieve robust and optimal constraints, which could lead to the discovery of new physics from large-scale structure observations.
	 
\matmethods{

    \subsection*{N-body Simulation and Lagrangian Perturbation Theory}

        Current cosmological $N$-body simulations model structure formation in the Universe by evolving a large number ($10^6 \sim 10^{12}$) massive dark matter particles that interact with each other only via gravity. Since our Universe begins with an almost uniform and Gaussian initial condition, the simulations start with particles only slightly perturbed from a uniform configuration, here a grid. For each particle, this initial small displacement $\mathbf{\Psi}$ from its grid location $\mathbf{q}$ can be computed accurately with the LPT. The particle position is then $\mathbf{x} = \mathbf{q} + \mathbf{\Psi}(\mathbf{q})$. The leading order LPT (ZA) predicts linear particle motion $\mathbf{\Psi}_\mathrm{ZA}(\mathbf{q}, z) = D(z) / D(z_0) \mathbf{\Psi}_\mathrm{ZA}(\mathbf{q}, z_0)$, where the growth function $D$ is a function of redshift $z$. As a result, the ZA velocity is linearly related to the displacement $\mathbf{v}_\mathrm{ZA}(\mathbf{q}, z) = a H(z) f(z) \mathbf{\Psi}_\mathrm{ZA}$, where $a=1/(1+z)$ is the expansion scale factor, $H=\dee\log a/\dee t$ is the Hubble expansion rate, and $f=\dee\log D/\dee\log a$ is known as the growth rate.

        Structures collapse under gravity and cluster hierarchically in the Universe. On small scales, particle motion becomes nonperturbative, invalidating ZA and higher order LPT. This process becomes so nonlinear that it can only be modeled accurately by expensive N-body simulations via time integration of thousands of time steps. On the other hand, on large scales the Universe remains relatively uniform, and perturbation theories still apply. This motivates us to build a model to emulate the small-scale structure formation in the N-body simulations, while retaining the ZA predictions on large scales. CNNs integrate well into this framework, since they excel at local textures but lack a global view beyond their finite receptive fields. Therefore, our goal is to train CNN models to nonlinearize the linear ZA inputs to make predictions comparable in accuarcey to N-body simulations.

        For benchmarks, we compare the NN predictions to alternative fast simulation models. Fast approximate simulations~\cite{Tassev2013,Feng:2016yqz} save computation time by only integrating tens of time steps, thus are less accurate than the full $N$-body simulations. Here we compare the NN model to the popular method COLA (COmoving Lagrangian Acceleration)~\cite{Tassev2013}, as implemented in \texttt{L-PICOLA}~\cite{Howlett2015}. COLA solves the particle motions relative to the second order LPT trajectory. We run \texttt{L-PICOLA} with the same settings as the full $N$-body simulations, but for only 10 time steps.

    \subsection*{Physical Considerations}

        As explained above, we want the NN to keep the ZA predictions ($\mathbf\Psi_\mathrm{ZA}$ and $\mathbf v_\mathrm{ZA}$ at the target redshift) on large scales, where perturbation theories are valid and accurate. This accounts for the nonlocal gravitational interactions beyond the local views of CNNs. The CNN models can then take the linear inputs and form the appropriate small-scale textures. This is most easily implemented by a global residual operation, in which the CNN predicts the difference between the N-body and ZA motions.

        The design with global residual connection has other advantages too. One is that the Galilean symmetry is approximately preserved by effective data augmentations. This is because even though the ZA and N-body particle motions may be affected by large-scale shifts in displacements or velocities, e.g., by adding a global constant vector, their difference is not. The other advantage is that the large-scale dependence on cosmological parameters is automatically accounted for by the ZA inputs, so the NN emulators only need to model the small-scale cosmology modulation, as explained below. Therefore we train two separate networks for displacement and velocity, respectively, whose inputs are linearly related.

        By construction CNNs should preserve the translational symmetry if they are fully convolutional. However, this is broken in most cases by the nonperiodic paddings commonly adopted in computer vision tasks. Our models fully preserve the translational symmetry using the periodic boundary condition of the N-body simulation volume. In addition, they approximately preserve the rotational symmetry of the cubic geometry by data augmentations that rotate and reflect the input and output fields as in Ref.~\cite{He:2018ggn}.

    \subsection*{Network and Training}

        Before applying CNN models to our problem, we need to prepare the data in an image-like format, which is straightforward thanks to the uniform initial condition of the Universe. Both the ZA inputs and N-body target are functions of the Lagrangian or initial positions, $\mathbf{q}$, which form a uniform grid. Therefore the displacements form a 3D image, with 3 channels being the Cartesian components of the particle displacements. Likewise, we format the velocity fields in a similar way.

        For both displacement and velocity, we adopt a simple U-Net/V-Net~\cite{UNet,VNet} type architecture similar to that in Ref.~\cite{AlvesdeOliveira:2020yix}. It works on 4 levels of resolution connected in a ``V'' shape first by 3 downsampling layers and then by 3 upsampling layers, by stride-2 $2^3$ convolutions and stride-\nicefrac12 $2^3$ transposed convolutions, respectively. Blocks of 2 $3^3$ convolutions connect the input, the resampling, and the output layers. Similar to V-Net, a residual connection \citep[ResNet,][]{He2016}, here a $1^3$ convolution instead of identity, is added over each block. We use a batch normalization layer after every convolution except the first one and the last two, and a leaky ReLU activation (with negative slope 0.01) after each batch normalization, as well as the first and the second to last convolution layers. As in the original ResNet architecture, the second or last activation in each residual block applies after the addition. And as in U-Net, the inputs to each of the downsampling layers are concatenated to the outputs of the upsampling layers of the same resolution, at the top 3 resolution levels. All layers have 64 channels, except the input and the output (3), and those after concatenations (128). Finally, an important difference from the original U-Net architecture is that we add the input displacement or velocity fields directly to the output. Therefore the network is effectively learning the corrections from linear to nonlinear motions, as we have mentioned above.

        With the Lagrangian description, the simplest loss functions for training one can formulate aim to minimize the residuals in the displacements. Ref.~\cite{AlvesdeOliveira:2020yix} and \cite{Li:2020vor} have shown that combining that with a loss of Eulerian quantities can greatly improve the performance of the trained model in the Eulerian space. For the displacement model, we can compute the particle positions and thus their (Eulerian) density distribution, described by the overdensity $\delta(\mathbf{x}) \equiv n(\mathbf{x}) / \bar{n} - 1$, where $n(\mathbf{x})$ is the particle number in voxel $\mathbf{x}$ and $\bar{n}$ is its mean value. Specifically, we use the CIC, or trilinear scheme to assign particles to the voxels. The total loss for our displacement network is $\log L_\delta + \lambda \log L_{\mathbf{\Psi}}$, with $L_\delta$ and $L_{\mathbf{\Psi}}$ being the mean squared error (MSE) losses on $n(\mathbf{x})$ and $\mathbf{\Psi}(\mathbf{q})$, respectively. Combining the two losses with logarithm allows us we to ignore their absolute magnitudes and trade between the relative changes. The parameter $\lambda$ is a weight on this trade-off, and we find setting $\lambda = 3$ for the displacement network achieves faster training. For the velocity network, we use a loss of $\log L_{\mathbf v(\mathbf q)} + \log L_{\mathbf v(\mathbf x)} + \log L_{{\mathbf v}^2(\mathbf x)}$. The first term is the MSE loss of the particle velocities (in Lagrangian space thus the $\mathbf q$ argument). The second and the last terms are defined by
        \begin{equation}
            L_{{\mathbf v}^n(\mathbf x)} = \int m \Bigl\|
            \Delta \bigotimes_{i=1}^n \mathbf v
            \Bigr\|_2^2 \,\dee{\mathbf x},
        \end{equation}
        where $m$ is the particle mass and $\otimes$ forms tensor products of the particle velocity and $\Delta$ takes the residual from predicted to N-body results. For $n=1$, this is simply the loss on the momentum fields, and for $n=2$ the loss minimizes the residual of the mass weighted velocity dispersions. We find the $n=2$ loss crucial for correctly predicting the redshift space distortion, especially the Fingers of God effect, as shown in \autoref{fig:stk}.

        The most important feature of our models is the ability to incorporate the dependence on cosmological parameters. The late-time structure evolution is sensitive to the expansion history of the Universe, i.e., to the expansion rate $H_0 = 100h \, \mathrm{km}/\mathrm{s}/\mpc$ and the matter density parameter $\om$. The former has been accounted for by using proper units that include $h$, thus we only need to include $\om$ as an input to our networks. We achieve this following StyleGAN2 \cite{StyleGAN2}, in which style parameters are incorporated by weight modulations and demodulations on the convolution kernels. In the modulation phase, the style parameters (here $\om$) are first affine transformed to the same dimension as the number of input convolution channels, and are then multiplied to the kernel of each channel, respectively. In the demodulation phase, the convolution kernels are renormalized by their fan-ins, so that propagating data do not explode or vanish. We refer readers to Eq.~(1--3) in Ref.~\cite{StyleGAN2} for more details.

        Limited by the size of GPU memory, an entire simulation field ($512^3$) cannot be fed at once to the network and has to be cropped into smaller chunks of size $128^3$. To preserve the translational equivariance, we keep no paddings in the convolutions, resulting in smaller network outputs than their inputs, and compensate this by periodically padding 48 voxels per side to the input fields. We train with a batch size of 16 distributed on 16 Nvidia V100 GPUs, and use the Adam optimizer~\cite{Kingma2014} with learning rate $0.0004$, and hyperparameters $\beta_1 = 0.9$, $\beta_2 = 0.999$. We reduce the learning to 10\% when the loss does not improve for 3 epochs.

        Finally, it's worth mentioning that the total data size amounts to 12 terabytes. This poses technical challenges because training the networks requires repeatedly loading the whole data set to the GPU nodes. If the data sampling and loading are not carefully designed the GPUs can easily outrun the cluster IO and be starved for data. We solve this and implement efficient training in \texttt{map2map}.

    \subsection*{Data Availability}

        Our trained model parameters are available at \url{github.com/dsjamieson/map2map_emu}. We train and test the neural networks with \texttt{map2map} (\url{github.com/eelregit/map2map}). \texttt{map2map} is a neural network framework for field-to-field emulators, based on \texttt{PyTorch}\cite{pytorch}. It implements the aforementioned mechanisms to fully preserve translational symmetry including the boundary conditions, and to do rotational data augmentation, for problems of arbitrary dimensions. It can also optimize the data loading IO at training time, so that the GPUs are not starved of data due to their high performance.
}

\showmatmethods{}

\acknow{The Flatiron Institute is supported by the Simons Foundation.}

\showacknow{}

\bibliography{neural_network_emulator}

\end{document}